
\def\a{\alpha}\def\b{\beta}\def\c{\chi}\def\d{\delta}\def\e{\epsilon}
\def\f{\phi}\def\h{\theta}
\def\l{\lambda}\def\m{\mu}\def\n{\nu}\def\o{\omega}\def
\p{\pi}\def\q{\psi}\def\r{\rho}\def\t{\tau}
\def\y{\eta}\def\x{\xi}

\def\H{\Theta}\def\L{\Lambda}
\def\O{\Omega}\def\Q{\Psi}\def\S{\Sigma}

\def\lie{{\cal L}}\def\de{\partial}\def\na{\nabla}\def\per{\times}
\def\inf{\infty}\def\id{\equiv}\def\mo{{-1}}\def\ha{{1\over 2}}
\def\di{{\rm d}}

\def\gmn{g_{\m\n}}\def\ghmn{\hat g_{\m\n}}
\def\dx{\int d^2x\ \sqrt g\ }\def\ds{ds^2=}
\def\dhx{\int d^2x\ \sqrt{\hat g}\ }

\def\af{asymptotically flat }\def\hd{higher derivative }\def\st{spacetime }
\def\fe{field equations }\def\bh{black hole }
\def\tran{transformations }\def\coo{coordinates }
\def\bg{background}\def\bhs{black holes }

\def\sch{Schwarzschild }\def\ads{anti-de Sitter }

\def\section#1{\bigskip\noindent{\bf#1}\smallskip}

\magnification=1200
\def\ef{e^{-2\f}}\def\kk{{k\over k-1}}\def\ehf{e^{-2h\f}}

\def\ep{\e_{ab}}\def\ab{^a_{\ b}}\def\ba{_a^{\ b}}\def\em{{\cal M}}
\def\epg{extended Poincar\'e group }
\def\nah{\hat\nabla}\def\Lh{{\L\over 2}}
\def\hp{{h+1}}\def\hi{{1\over h+1}}\def\lx{(\l x)^{h+1}-c}\def\Lk{{\L\over 4}}
\def\ehl{e^{h\l y}}\def\el{e^{-\l y}}

{\nopagenumbers
November 1994\hfill INFNCA-TH-94-28
\vskip80pt
\centerline{\bf Black holes in generalized dilaton gravity in two dimensions}
\vskip40pt
\centerline{\bf S. Mignemi\footnote{$^\dagger$}{\rm
e-mail: mignemi@cagliari.infn.it}}
\vskip20pt
\centerline{Dipartimento di Scienze Fisiche, Universit\`a di Cagliari}
\centerline{via Ospedale 72, 09100 Cagliari, Italy}
\centerline{and}
\centerline{INFN, Sezione di Cagliari}
\vskip80pt
\centerline{\bf ABSTRACT}
{\noindent We consider two-dimensional dilaton-gravity theories with a
generic exponential potential for the dilaton, and obtain the most general
black hole solutions in the \sch form. We discuss their geometrical and
thermodynamical properties. We also study these models from the point of
view of gauge theories of the extended Poincar\'e group and show that they
can be considered as gauge theories with broken symmetry. Finally, we examine
the theory in a hamiltonian formalism and discuss its quantization and its
symmetries.}
\vfil\eject}
\section{1. Introduction}
In this paper we study the \bh solutions of a class of two-dimensional
gravity-scalar theories with generic power-law scalar potential.
As is well-known, in two dimensions the Einstein-Hilbert action is
a total derivative and hence cannot be used to construct a 2-dimensional
version of general relativity. Only relatively recently it was realized
that an action for two-dimensional gravity can be written down if one
introduces a scalar field (whose logarithm is called dilaton in string
language), which is coupled non-minimally to gravity [1].
Indeed, a full class of lagrangians
can be written down, by varying the form of the kinetic and potential
terms for the scalar [2]. These can be reduced, via field redefinitions and
conformal transformations of the metric to a standard form, with vanishing
scalar kinetic term and arbitrary potential [2,3]. In particular, two special
cases have been extensively studied: the Jackiw-Teitelboim (JT) theory [1],
and the string theory action [4], which correspond to linear and flat
potential respectively [5]. In this paper we discuss the more general case of
power-law potentials, which interpolate between the two.\footnote{$^\dagger$}
{In terms of the dilaton, the potentials are of course exponential.}
These potentials
are especially interesting because the corresponding actions are equivalent
to higher-derivative actions containing powers of the Ricci scalar [6]:
higher-derivative theories have indeed been proposed as an alternative way
to model gravity in two dimensions [7]. Moreover, the introduction of a
potential term in dilaton-gravity theories is useful in the context of
string-generated models, since it may provide a mass for the dilaton.
With this motivation, a model similar to ours has been studied in
four dimensions in [8].

A different formulation of
two-dimensional gravity is given by the gauge formalism [9,10]. The JT and
string-like models can in fact be written down as topological gauge theories
of the
anti-de Sitter and extended Poincar\'e group respectively. It has been shown
[6] that also the models considered here can be formulated in the framework
of extended Poincar\'e theory, if some symmetry-breaking terms are added to
the gauge-invariant lagrangian. In this paper we proceed further and discuss
the quantization of the generalized actions in this framework.
We also discuss an alternative formulation of the symmetries of the theory,
involving a non-linear generalization of the de Sitter algebra.

The paper is organized as follows: in section 2 we review the model and
discuss the physical properties of its exact \bh solutions. In section 3,
we discuss the conformally related theory formulated in terms of the metric
which is relevant for string theory. In particular, we discuss the \bh
solutions in a \sch gauge, where they exhibit more clearly their physical
properties than in the conformal gauge adopted in [6].
In section 4 we reformulate the model in the gauge formalism and comment
about its quantization and its invariance properties.

\section{2. The canonical action}
We consider the 2-dimensional action:
$$S=\dx (\y R+\L\y^h)\eqno(1)$$
For $h=1$, it reduces to that of the JT theory [1], while for $h=0$, it is
conformally related to the low-energy effective string action [4].
It has been shown in [6] that the action (1) is also related to the
\hd action
$$S=\dx R^k\eqno(2)$$
by the field redefinition $\y=kR^{k-1}$, where $h=k/(k-1)$ and
$\L=(1-k)k^{-\kk}$, $k\ne 0, 1$.

Variation of the action (1) yields the \fe
$$R=-\L h\y^{h-1}\eqno(3a)$$
$$(\na_\m\na_\n-\gmn\na^2)\y+{\L\over 2}\gmn\y^h=0\eqno(3b)$$
Let us consider the case $\L>0$. The general static solutions of (3) can be
easily found in the \sch gauge $\ds -Adt^2+A^\mo dx^2$ [6,7]:
for $h\ne-1$, one has
$$A=\hi[\lx]\qquad\qquad\y=\l x\eqno(4)$$
where $\l=\sqrt\L$  and $c$ is an
integration constant.
These solutions are related to those found in the conformal gauge in [6],
by simply taking the scalar field $\y$ as the spacelike coordinate.
The curvature of the metric (4) is given by:
$$R=-A''=-h\l^{h+1}x^{h-1}\eqno(5)$$

For $c=0$, the metric is regular (and flat) at $x=\pm\inf$ iff $h<1$, and is
singular at $x=0$. The opposite happens if $h>1$. The two special cases $h=0$
and $h=1$ correspond respectively to flat and \ads space. If $h<1$, $x=\inf$
is at infinite spatial distance and the singularity at finite distance. If
$h>1$, the singularity at $x=\inf$ is at finite distance, while $x=0$ is at
infinite distance. The two cases present therefore essentially the same
physical
behaviour, with the r\^ole of $x=0$ and $x=\inf$ interchanged. The flat end
of the \st at
spatial infinity is not \af in the usual sense, since the metric is not
constant there, but is more similar to a horizon. The
$c=0$ configuration can alternatively be written in the unitary gauge as:
$$\ds dr^2-r^{-2{h+1\over h-1}}dt^2\eqno(6)$$

Let us now consider the case of nonvanishing $c$. It is easy to see that in
general ($h\ne 0,1$), for $c<0$ a naked singularity is present at $x=0$.
For $c>0$, instead, a horizon is placed at
$\l x=c^{1/(h+1)}$. The metric describes in this case a regular, \af\bh\st.
In the special case $h=1$, however, the solution describes the constant
curvature \ads regular \bh discussed in [11]. For $h=0$, instead, one has
flat space in non-standard \coo.

The ADM mass $M$ of the black hole solutions can be most easily calculated by
means of the formalism of Mann [12]: in his notations
$$M={1\over 2\l}\left[\l^2\y^{h+1}-(\nabla\y)^2\right]$$
Substituting the values for the $c>0$ solutions and subtracting the
contribution of the $c=0$ \bg, one obtains:
$$M={\l\over 2(h+1)}c\eqno(7)$$

The temperature of the \bh at the horizon is given by the inverse periodicity
of the regular euclidean solution:
$$T={\l\over 4\p}c^{h\over h+1}\eqno(8)$$
{}From the formulae above, it is clear that for $h>0$ or $h<-1$, $T$ is an
incresing function of the mass, and vanishes for $M=0$. The ground state
for the Hawking evaporation process is given in these cases by the $c=0$
solution. For $-1<h<0$, instead, the behaviour  of $T$ is more similar to
that of the \sch solution in general relativity, with $T$ divergent at $M=0$
(see fig. 1a). For $h=0$, $T$ is independent of the mass.

The entropy $S$ can be obtained by integrating the thermodynamical relation
$dS=T^\mo dM$, yielding:
$$S=2\p c^\hi\eqno(9)$$
For $h>-1$, the entropy vanishes at $M=0$ and increases with the mass, while
for $h<-1$ it is a decreasing function of $M$, which diverges for $M=0$
(see fig. 1b).
If $h>0$ or $h<-1$, the entropy increases when a \bh splits into smaller
holes, while it decreases if $-1<h<0$. A relation between $S$ and
$T$ which is valid for all these models is
$$ST={\l c\over 2}=(h+1)M\eqno(10)$$

Finally, we consider the special case $h=-1$. The solution of the field
equations is in this case:
$$A=\ln(\l x)-c\qquad\qquad\y=\l x\eqno(11)$$
Proceeding as before, one can calculate mass, temperature and entropy of the
solution. They are given by:
$$M={\l c\over 2}\qquad\qquad T={\l\over 4\p}e^{-c}\qquad\qquad S=2\p e^c
\eqno(12)$$
The temperature is thus finite for $M=0$ and decreases for incraesing $M$.

The equations of motion of a particle moving in the metric (4) can be easily
obtained by varying the line element. The result is:
$${dt\over d\t}=E\sqrt{h+1\over\lx}\qquad\qquad{dx\over d\t}=\sqrt{E^2-
{\e\over h+1}[\lx]}\eqno(13)$$
where $\t$ is the proper time and $\e=1, 0$ for massive (resp. massless)
particles. While for massless particles one simply has $x=E\t$, massive
particles experience a gravitational potential whose shape depends on $h$ (see
fig. 2). If $h<1$ the particles are attracted towards the singularity, while
if $h>1$ they are repelled (we recall that for $h>1$ the singularity is at
infinity). If $h>|1|$, the potential diverges at the singularity as in the
\sch case, while for $h<|1|$ it is regular. The singularity is always reached
in a finite proper time, while spatial infinity requires an infinite amount of
proper time to be achieved.

To conclude this section, we make some comments about the case of negative
$\L$. The metric is given in this case by $A=c-\l x^{h+1}$, with $\l=\sqrt
{-\L}$. A cosmological horizon is located at $\l x=c^{1/h+1}$, and the
solution is regular at the origin iff $h>1$ (otherwise a naked singularity
is present). In this case the solutions are qualitatively similar to the
2-dimensional de Sitter \st.

We also notice that it is possible to generalize the solutions (3) to the case
of a scalar potential of the form $V(\y)=\S_i\L_i\y^{h_i}$. In this case the
solutions are given by
$$A=\sum_i {\L_i\over h_i+1}x^{h_i+1}-c\qquad\qquad\y=x$$
These solutions are of course more complicated than (3): in particular,
they may possess multiple horizons.

\section{3. The stringy action}
A conformal transformation of the metric $\ghmn=e^{2\f} \gmn$, with $\ef=\y$,
puts the action (1) in the form [6]:
$$S=\dhx\ef[\hat R+4(\nah\f)^2+\L\ehf]\eqno(14)$$
This is a generalization of the low-energy string action of [4], which is
obtained in the limit $h=0$. The field equations
$$4\nah^2\f-4(\nah\f)^2+(1+h)\L\ehf+\hat R=0\eqno(15a)$$
$$-2\nah_\m\nah_\n\f-\ghmn [-2\nah^2\f+2(\nah\f)^2-{\L\over 2}\ehf]=0
\eqno(15b)$$
are solved in the \sch gauge $\ds -Adt^2+A^\mo dx^2$, by ($\l^2=\L>0$):
$$\hat A=\hi(\ehl-c\el)\qquad\qquad\hat\f=-{\l\over 2}y\eqno(16)$$
These solutions can be obtained from those discussed in the conformal gauge
in [6], by choosing the dilaton as the spatial coordinate. Alternatively, they
can be obtained from (4), performing the conformal \tran $\ghmn=e^{2\f} \gmn$
and then defining the new variable $\l y=\ln (\l x)$.
The curvature of the solution is given by
$$\hat R={\l^2\over h+1}(h^2\ehl-c\el)\eqno(17)$$

If $c=0$, the metric is regular at $y=-\inf$ and singular at $y=\inf$ for
positive $h$ (the opposite for negative $h$). For $h=0$, one has flat space.
The singularity is at finite spatial distance, while the regular spacetime
end is at infinite distance.  The metric is flat at infinity, but not \af
in the usual sense, since $y=\inf$ is a null line. In the unitary gauge,
the $c=0$ solutions can also be written for nonvasnishing $h$ as
$$\ds d\r^2-\r^{-2}dt^2\eqno(18)$$

If $c\ne0$ one must distinguish two cases: for positive $h$ the metric is
singular at both $y=\pm\inf$. If instead $h<0$, the metric is regular at
$y=\inf$ and diverges at $y=-\inf$. Moreover, a horizon is present at
$e^{(h+1)\l y}=c$, for positive $c$. In this case, the metric describes a \bh
with singularity at $y=-\inf$ and \af at positive infinity. If $c<0$, instead,
a naked singularity is present at finite distance. For $h=0$, one recovers
the MSW solutions of the string lagrangian [4].

The mass of the \bh solutions can be obtained as before, following [12]: in
the present case
$$M={1\over 2\l}[\l^2e^{-2(h+1)\f}-4\ef(\nabla\f)^2]$$
Subtracting the background value ($c=0$), one has:
$$M={\l\over2(h+1)}c\eqno(19)$$
which coincides with (7). Analogously, for the temperature and the entropy
of the solutions, one obtains the results (8) and (9).

In the special case $h=-1$, the solution is given by
$$A=(\l y-c)e^{-\l y}\qquad\qquad\f=-{\l\over 2}y\eqno(20)$$
Again, its thermodynamical parameters coincide with those of the
corresponding solution (11), which are listed in (12).

The motion of a particle in the metrics (16) is obtained by solving the
differential equations:
$${dt\over d\t}=E\sqrt{h+1\over\ehl-c\el}\qquad\qquad{dy\over d\t}=
\sqrt{E^2-{\e\over h+1}(\ehl-c\el)}\eqno(21)$$
where $E$ is the energy of the particle and $\t$ its proper time. Again,
massless particles move along $y=E\t$, while massive particles experience the
potential depicted in fig. 3. For $h<0$, the potential is attracive at
short distances and repulsive at large distances. A particle coming from
infinity must therefore have energy $E^2>(-c/h)^{h/(h+1)}$ in order to reach
the singularity.
For positive $h$, instead, the particle is always repelled from the
singularity (placed at $y=\inf$).

The solutions with $\L<0$ are given by $A=c\el-\ehl$, $\l=\sqrt{-\L}$. They
have a cosmological horizon at $e^{(h+1)\l y}=c$, but all are singular at
at $y=-\inf$ and hence do not seem to be physically relevant.

Also in this case the solutions (16) can be generalized to potentials of the
form $V(\f)=\S_i\L_ie^{-2h_i\f}$. The solutions are given now by:
$$A=\sum_i{\L_i\over h_i+1} e^{h_i x}-ce^{-x}\qquad\qquad\f=-{x\over 2}$$
The properties of these solutions depend of course on the specific form of
the coefficients $\L_i$ and $h_i$.

\section{4. The gauge formulation}
In some circumstances, two-dimensional dilaton-gravity models can be
considered as gauge theories.
In particular, it has been shown [9], that the JT model can be
formulated as a gauge theory of the 2-dimensional (anti)-de Sitter group.
Similarly, the string-inspired model ($h=0$), can be thought of as a gauge
theory of the 2-dimensional extended Poincar\'e group [10]. In this section
we show that also our generalized models can be formulated as gauge theories
of the \epg, if a symmetry-breaking term is added to the lagrangian [6].

Consider the 2-dimensional extended Poincar\'e algebra [10]:
$$[P_a, J]=\e\ab P_b\qquad\qquad [P_a, P_b]=\ep I\qquad\qquad
[P_a, I]=[J, I]=0$$
and the corresponding gauge field:
$$A=e^aP_a+\o J+a I\eqno(22)$$
where $e^a$ is the zweibein and $\o$ is the spin connection.
The field strength is given by
$$F=dA+A^2=P_aT^a+Jd\o +I\left(da+\ha\ep e^ae^b\right)\eqno(23)$$
with the torsion $T^a\id de^a+\e\ab\o e^b$.

According to [8], the fields transform under the gauge \tran generated by
$\H=\h^a P_a+\a J+\b I$, as:
$$\eqalign{&e^a\to(\em^\mo)\ab(e^b+\e^b_{\ c}\h^c+d\h^b)\cr
&\o\to\o+d\a\cr
&a\to a-\h^a\ep e^b-\ha\h^2\o+d\b+\ha d\h^a\ep\h^b}\eqno(24)$$
where $\em\ab=\delta\ab\cosh\a+\e\ab\sinh\a$.

One can now define the gauge multiplet of scalar fields $\y_A=(\y_a,\y_2,
\y_3)$, which permits to construct the topological lagrangian:
$$\ha\lie_1=\S \y_A F^A=\y_a T^a+\y_2 d\o+\y_3(da+\ha\ep e^ae^b)\eqno(25)$$
invariant under the extended Poincar\'e group. To this we add a
symmetry-breaking term, which is invariant only with respect to the
subgroup $U(1)\per U(1)$, generated by $J$ and $I$:
$$\ha\lie_2=\c_1da+\c_2\left(\y_3-\Lh\y_2^h\right)\eqno(26)$$
The lagrangian multipliers $\c_i$ enforce the constraints:
$$da=0\qquad\qquad\y_3=\Lh\y_2^h\eqno(27)$$
Solving these constraints, one obtains the lagrangian
$$\ha\lie=\y_a T^a+\y_2 d\o+\Lk\y_2^h\ep e^ae^b\eqno(28)$$
For $h=1$, the action (28) coincides with that of \ads gravity [9].
For $h=0$, instead, the constraints (27) reproduce the explicit
solution [10] for $a$ and $\y_3$
of the field equations of the unconstrained lagrangian (25).

The field equation stemming from (28) are
$$\eqalign{&T^a=0\cr
&R+h\L\y_2^{h-1}=0\cr
&\di\y_a+\e_{\ a}^b\o\y_b+\Lh\y_2^h\ep e^b=0\cr
&\di\y_2+\y_a\e^a_{\ b}e^b=0}\eqno(29)$$
The first equation implies the vanishing of the torsion and hence the usual
relation between the spin connection and the zweibein, while the second
coincides with (3a), provided one identifies $\y_2$ with $\y$.
Finally, the last two equation, combined, yield (3b).
The solutions of the field equations are therefore given by (4), and hence
the nonvanishing components of the zweibein and the connection are:
$$e^0_t=\sqrt{\lx\over h+1}=(e^1_x)^\mo\eqno(30)$$
$$\o_t=\ha\l^{h+1}x^h\qquad\qquad\o_x=0$$
Moreover, $\y_2$ can be identified with the scalar field $\y$ in (1), yielding
$$\y_2=\l x\eqno(31)$$
Finally, one can obtain $\y_a$ by solving the third equation (29). The
result is:
$$\y_0=-\l\sqrt{\lx\over h+1}\qquad\qquad\y_1=0\eqno(32)$$

One can now evaluate the mass of the solution as in [13]; after subtracting
as usual the vacuum contribution, one gets:
$$M=\y_a e^a_t+\y_2\o_t\Big|^\inf={\l\over 2(h+1)}c\eqno(33)$$
which is the same result (7)
obtained before in the geometric approach.

In order to quantize the model, it is interesting to consider the hamiltonian
formulation of the equations of motion. The lagrangian (27) can be written,
after integration by parts:
$$\ha\lie=\y_a\dot e^a_1+\y_2\dot\o_1+e^a_0(\y_a'+\e\ba\y_b\o_1+\Lh\y_2^h
\ep e_1^b)+\o_0(\y_2'+\y_a\e\ab e^b_1)\eqno(34)$$
where a dot denotes time derivative and a prime spatial derivative.

The lagrangian (34) has a canonical structure, with coordinates $(e^a_x,
\o_x)$, conjugate momenta $(\y_a,\y_2)$ and Lagrange multipliers $(e^a_t,
\o_t)$ enforcing the constraints:
$$G_a=\y_a'+\e\ab\y_b\o_x+\Lh\y_2^h\ep e_x^b=0$$
$$G_2=\y_2'+\y_a\ep e^b_x=0\eqno(35)$$
which imply the conservation of the quantity $\y^a\y_a-{\L\over\hp}\y_2^\hp$,
whose value coincides with the mass of the solution.
The algebra of constraints is given by
$$\{G_a,G_2\}=\e\ba G_b\qquad \{G_a,G_b\}=\ep\Lh(\y_2^h)'G_2\eqno(36)$$
where the structure constants are functions of the fields. The
algebra looks like a deformation of the SO(2,1) anti-de Sitter algebra and
corresponds to a non-linear local symmetry of the action (28), generated by
the infinitesimal transformations:
$$\d e^a=d\x^a+\e\ab(\x^b\o-\x^2e^b)\qquad\qquad
\d\o=d\x^2-\Lh h\y_2^\hp\ep\x^a e^b$$
$$\d\y_a=\e\ba\left(\Lh\x_b\y_2^h+\x^2\y_b\right)\qquad\qquad
\d\y_2=\e\ba\x^a\y_b$$

The quantization can now be straightforwardly performed by replacing the
Poisson brackets with commutators and imposing the Gauss law on the physical
states, with an appropriate operator ordering. In a Schr\"odinger
representation
$e^a\to i{d\over d\y_a}$, $\o\to i{d\over d\y_2}$, the constraint equations
become:
$$\left(\y_a'+i\e\ba\y_b{\de\over\de\y_2}+i\Lk\ep\y_2^h{\de\over\de\y_b}
\right)\Q(\y_a,\y_2)=0\eqno(37a)$$
$$\left(\y_2'+i\e\ab\y_a{\de\over\de\y_b}\right)\Q(\y_a,\y_2)=0
\eqno(37b)$$
The solution of (37) is analogous to that given in [14] for the $h=1$ case:
$$\Q=\d\left(\left[\y^a\y_a-{\L\over\hp}\y_2^\hp\right]'\right)e^{i\O}\q_
\L(M)\eqno(38)$$
where $M$ is the mass (7) of the solution and
$$\O=\int dx{\y_2\e^{ab}\y_a\y_b'\over\y^c\y_c}\eqno(39)$$
The physical states are therefore classified by the mass $M$.
We shall not consider further the quantization of the model, which has also
been studied in the geometrical formulation in [15].

\section{5. Final remarks}
We have studied the geometrical and thermodynamical properties of
two-dimensional gravity-scalar \bhs in two different gauges corresponding
to the canonical and string metric.
They possess identical thermodynamical properties, but have different
geometries. Which of the two gauges is relevant for physics depends of
course on the specific form of the matter coupling, which has not been
discussed here.

We also have investigated the gauge formulation of the theory. Our models
can be obtained by adding symmetry-breaking terms to the \epg theory.
This symmetry breaking could also be obtained dynamically: preliminary
results indicate that it can indeed be obtained at the expense of introducing
non-vanishing torsion into the theory.

Alternatively, we have shown that the action of the theory is invariant under
a non-linear deformation of the anti-de Sitter group, with field-dependent
structure constants. It would be interesting to investigate in more detail
the mathematical structure of this symmetry.
\bigskip{\noindent\bf Acknowledgements}

I wish to thank the Departament de F\'isica i Enginyeria Nuclear of the
Universitat Polit\'ecnica de Catalunya
and especially Ll. Ametller for kind hospitality during the first stages
of the preparation of this work.

\bigskip
\centerline{\bf REFERENCES}
\smallskip
\def\PL#1{Phys.\ Lett.\ {\bf#1}}
\def\PRL#1{Phys.\ Rev.\ Lett.\ {\bf#1}}
\def\PR#1{Phys.\ Rev.\ {\bf#1}}\def\CQG#1{Class.\ Quantum Grav.\ {\bf#1}}
\def\NP#1{Nucl.\ Phys.\ {\bf#1}}
\def\JMP#1{J.\ Math.\ Phys.\ {\bf#1}}

\def\MPL#1{Mod.\ Phys.\ Lett.\ {\bf #1}}

\halign{#\quad&#\hfil\cr
[1]& C. Teitelboim, in {\sl Quantum Theory of gravity}, S.M. Christensen,
 ed. (Adam Hilger,\cr & Bristol, 1984); R. Jackiw, {\sl ibidem};\cr
[2]&T. Banks and O'Loughlin, \NP{B362}, 649 (1991);\cr
&J.G. Russo and A.A. Tseytlin, \NP{B382}, 259 (1992);\cr
[3]&S. Mignemi and H.-J. Schmidt, preprint INFNCA-TH-94-24;\cr
[4]&G. Mandal, A.M. Sengupta and S.R. Wadia, \MPL{A6}, 1685 (1991);\cr
&E. Witten, \PR{D44}, 314 (1991);\cr
[5]&R. Jackiw, Theor. Math. Phys. {\bf 9}, 404 (1992);\cr
[6]&S. Mignemi, \PR{D50}, in press;\cr
[7]&H.J. Schmidt, \JMP{32}, 1562 (1991);\cr
[8]&S. Poletti and D.L. Wiltshire, preprint ADP-94-2/M22;\cr
[9]&K. Isler and C.A. Trugenberger, \PRL{63}, 834 (1989);\cr
&A.H. Chamseddine and D. Wyler, \PL{B228}, 75 (1989);\cr
[10]&D. Cangemi and R. Jackiw, \PRL{69}, 233 (1992);\cr
[11]&D. Christensen and R.B. Mann, \CQG{9}, 1769(1992);\cr
&A. Ach\' ucarro and M.E. Ortiz, \PR{D48}, 3600 (1993);\cr
&J.P.L. Lemos and P.M. S\'a, \MPL{A9}, 771 (1994);\cr
&M. Cadoni and S. Mignemi, preprints INFNCA-TH-94-4 and INFNCA-TH-94-26;\cr
[12]&R.B. Mann, \PR{D47}, 4438 (1993);\cr
[13]&D. Bak, D. Cangemi and R. Jackiw, \PR{D49}, 5173 (1994);\cr
[14]&D. Cangemi and R. Jackiw, \PR{D50}, 3913 (1994);\cr
[15]&D. Louis-Martinez, J. Gegenberg and G. Kunstatter, \PL{B321}, 193 (1994).
\cr
}
\end